\documentclass[11pt]{article}
\usepackage{amsmath,amssymb,epsfig,bm}
\usepackage{mathrsfs}
\date{\empty}

\begin{document}

\title{\bf Gravitational-wave implications for structure formation: A second-order approach}

\author{Despoina Pazouli\footnote{Current address: DAMTP, CMS, University of Cambridge, Wilberforce Road, Cambridge CB3 0WA, UK.}\; and Christos G. Tsagas\\ {\small Section of Astrophysics, Astronomy and Mechanics, Department of Physics}\\ {\small Aristotle University of Thessaloniki, Thessaloniki 54124, Greece}}

\maketitle

\begin{abstract}
Gravitational waves are propagating undulations in the spacetime fabric, which interact very weakly with their environment. In cosmology, gravitational-wave distortions are produced by most of the inflationary scenarios and their anticipated detection should open a new window to the early universe. Motivated by the relative lack of studies on the potential implications of gravitational radiation for the large-scale structure of the universe, we consider its coupling to density perturbations during the post-recombination era. We do so by assuming an Einstein-de Sitter background cosmology and by employing a second-order perturbation study. At this perturbative level and on superhorizon scales, we find that gravitational radiation adds a distinct and faster growing mode to the standard linear solution for the density contrast. Given the expected weakness of cosmological gravitational waves, however, the effect of the new mode is currently subdominant and it could start becoming noticeable only in the far future. Nevertheless, this still raises the intriguing possibility that the late-time evolution of large-scale density perturbations may be dictated by the long-range (the Weyl), rather than the local (the Ricci) component of the gravitational field.
\end{abstract}

\section{Introduction}\label{sI}
%%%%%%%%%%%%%%%%%%%%%%%%%%%%%%%%
Cosmological gravitational waves, namely pure-tensor (transverse-traceless) perturbations, are one of the prominent predictions made by the simplest and most popular models of inflation~\cite{RSV}. The detection of these distortions will provide the test-ground for the various inflationary models, while mapping the gravitational-wave background could lead to the phenomenological reconstruction of the inflaton.\footnote{Recently the first ever detection of astrophysical gravitational waves, emitted during the merging of two massive black holes, was announced by the LIGO Collaboration~\cite{Aetal}.} According to the standard scenario, primordial gravitational waves originate as subhorizon quantum fluctuations, which are stretched on supper-Hubble scales by the exponential expansion of the de Sitter phase. Once there, they ``freeze-out'' as classical Weyl-curvature distortions that interact very weakly with their environment. Because of that, the detection (direct or indirect) of a cosmological gravitational-wave background should provide us with valuable new information about the very early stages of our universe. An announcement of a possible indirect detection of such inflationary gravitational waves was made by the BICEP2 Collaboration in March 2014~\cite{Aetal1}. Although the confidence to this first-ever detection has now subsided~\cite{FHS}, the search for a primordial gravitational-wave background continues~\cite{Aetal2}.

Distortions in the long-range component of the gravitational field interact with those in the density of the matter, though typically not at the linear level. In the literature there already second-order studies looking into the effects of scalar density perturbation on gravitational waves (e.g.~see~\cite{BSTI} and references therein). Here we will consider the opposite, namely the implications of gravitational radiation for the evolution of density perturbations during the dust epoch. A related study, looking at the effects of relic gravitons on scalar curvature perturbations throughout the radiation and the dust eras, has also recently appeared in the literature~\cite{G}, following earlier quantum mechanical studies of the de Sitter phase~\cite{W}. It was shown in~\cite{G} that inflationary produced, large-scale, gravitational waves add a weak growing mode to the evolution of curvature perturbations during both the radiation and the dust eras. The effect of the new mode was found to depend on the ratio between the graviton energy-density and that of the dominant matter component~\cite{G}. Here, we arrive at analogous results by looking into the gravitational-wave effects on superhorizon-sized density perturbations during the dust era and by using considerably different analytical techniques.

Our study is non-perturbative, in the sense that it starts from the fully nonlinear expressions before reducing them to their linear and second-order limits (around a chosen background). This is achieved by employing the covariant 1+3 splitting of the Weyl tensor, in order to describe the gravitational waves, instead of perturbing the background metric. Assuming an Einstein-de Sitter background, we isolate the gravitational waves and use the transverse component of the shear tensor to describe these distortions at the linear level. Confining to superhorizon scales, we then look into the effects of the Weyl field on the evolution of overdensities/underdensities to second-order. We arrive at an inhomogeneous differential equation with source-terms due to the gravitational waves. Solving this equation we recover the standard growing and decaying modes of the linear analysis, plus a new gravitationally-induced mode that grows considerably faster than its first-order counterpart. Nevertheless, the expected weakness of cosmological gravitational waves means that the aforementioned additional mode is currently still subdominant and can only have a measurable effect in the far future. Put another way, our second-order study suggests that Weyl-curvature distortions could in principle dictate the evolution of large-scale density perturbations, although this will probably happen only at very late times. We should also note that, given the purely geometrical nature of the interaction under consideration, the matter component does not need to be necessarily baryonic. Therefore, our analysis and results apply to ordinary pressure-free dust and to non-baryonic Cold-Dark-Matter (CDM) as well.

During our analysis we also encounter the so-called ``gauge issue'', stemming from the fact that in cosmological perturbation studies we deal with two spacetimes. An idealistic one, which usually coincides with a Friedmann-Lemaitre-Robertson-Walker (FLRW) model, and a more realistic perturbed spacetime that is expected to describe the actual universe. To proceed one needs to establish an one-to-one mapping, namely a ``gauge'', between these two spacetimes and in many occasions the results depend on the chosen gauge. We bypass the gauge problem, by employing a new variable that is gauge-invariant to second order. The aforementioned quantity has the advantages of being covariantly defined, directly related to (scalar) density perturbations and very simple to construct. On the downside, to guarantee mathematically the gauge-invariance of our second-order results, we must confine to  the uniform component of the linear overdensities/underdensities. Physically, this assumption implies that we consider the average linear density contrast inside the associated overdensity/underdensity. The latter is a fairly common practice in nonlinear perturbation studies~\cite{P}.

We start with a general introduction to the 1+3 covariant formalism in section 2 and then proceed to discuss the basics of the Friedmannian cosmologies in section 3. In the next two sections (4 and 5) we consider the the linear evolution of gravitational waves and density perturbations respectively. The nonlinear interaction between these two physical entities is presented in section 6. There, we also address the gauge issue and provide an analytical solution for the second-order evolution of (scalar) density perturbations in the presence of gravitational-wave distortions. Finally, we discuss the implications of our results in section 7. For the interested reader, we have also included two appendices with additional technical details.

\section{The 1+3 covariant approach to gravitational %%%%%%%%%%%%%%%%%%%%%%%%%%%%%%%%%%%%%%%%%%%%%%%%%%%%
waves}\label{s1+3CF}
%%%%%%%%%%%%%%%%%%%%%
The 1+3 covariant approach (see~\cite{TCM,EMM} for recent reviews) describes physics by introducing a family of (fundamental) observers, which split the spacetime into time and 3-D space. Then, all variables, operators and equations decompose into their temporal and spatial parts.

\subsection{Kinematics}\label{ssKs}
%%%%%%%%%%%%%%%%%%%%%%%%%%%%%%%%%%%
Consider a family of observers `living' along worldlines with local coordinates $x^a=x^a(\tau)$, where $\tau$ is the observers' proper time. The (timelike) 4-velocity tangent to these worldlines is $u^a={\rm d}x^a/{\rm d}\tau$, so that $u^au_a=-1$. This 4-velocity vector defines the temporal direction, while the tensor $h_{ab}=g_{ab}+u_a u_b$ projects into the 3-dimensional hypersurfaces orthogonal to $u_a$~\cite{TCM,EMM}. The $u_a$-field and its tensor counterpart $h_{ab}$ facilitate a unique decomposition of any spacetime quantity, operator and equation into their timelike and spacelike components.

The motion of the fundamental observers is characterized by a set of irreducible kinematic quantities, which emerge from the 1+3 covariant decomposition of the 4-velocity gradient \cite{TCM,EMM}
\begin{equation}
\nabla_b u_a= \sigma_{ab}+ \omega_{ab}+ \frac{1}{3}\,\Theta h_{ab}- A_au_b\,.  \label{vdecomposition}
\end{equation}
In the above, $\sigma_{ab}={\rm D}_{\langle b}u_{a\rangle}$ and $\omega_{ab}= {\rm D}_{[b}u_{a]}$ are respectively the shear and the vorticity tensors, $\Theta={\rm D}^au_a$ is the volume scalar and $A_a$ is the 4-acceleration vector (with $\sigma_{ab}u^b=0=\omega_{ab}u^b=A_au^a$).\footnote{Round brackets denote symmetrization, square brackets imply antisymmetrization and angled brackets indicate the projected, symmetric and trace-free part of a tensor. In particular ${\rm D}_{\langle b}u_{a\rangle}={\rm D}_{(b}u_{a)}-({\rm D}^cu_c/3)h_{ab}$.} For our purposes the key variable is the shear tensor, which describes kinematic anisotropies and is directly related to gravitational waves in perturbed FLRW cosmologies. At the nonlinear level, the shear evolution is monitored by the propagation equation \cite{TCM,EMM}
\begin{equation}
\dot{\sigma}_{\langle ab\rangle}= -\frac{2}{3}\,\Theta\sigma_{ab} -\sigma_{c\langle a}\sigma^c{}_{b\rangle}- \omega_{\langle a}\omega_{b\rangle}+ {\rm D}_{\langle a}A_{b\rangle}+ A_{\langle a}A_{b\rangle}- E_{ab}+ \frac{1}{2}\,\pi_{ab}\,,  \label{shear}
\end{equation}
while it obeys the constraints
\begin{equation}
{\rm D}^b\sigma_{ab}= \frac{2}{3}\,{\rm D}_a\Theta+ \mbox{curl}\omega_a+ 2\varepsilon_{abc}A^b\omega^c- q_a \label{shearc}
\end{equation}
and
\begin{equation}
H_{ab}=\mbox{curl}\sigma_{ab}+{\rm D}_{\langle a}\omega_{b\rangle}+ 2A_{\langle a}\omega_{b\rangle}\,. \label{magneticeq}
\end{equation}
Note that overdots denote time derivatives (e.g. $\dot{\sigma}_{ab}=u^c\nabla_c\sigma_{ab}$). Also, $\omega_a=\varepsilon_{abc}\omega^{bc}/2$, ${\rm curl}\omega_a=\varepsilon_{abc}{\rm D}^b\omega^c$ and ${\rm curl}\sigma_{ab}=\varepsilon_{cd\langle a}{\rm D}^c\sigma^d{}_{b\rangle}$ by definition, with ${\rm D}_a=h_a{}^b\nabla_b$ representing the 3-dimensional covariant derivative operator. Finally, $q_a$ and $\pi_{ab}$ are the energy-flux vector and the anisotropic pressure of the matter respectively (with $\pi_{ab}=\pi_{\langle ab\rangle}$ and $q_au^a=0=\pi_{ab}u^b$), while $E_{ab}$ and $H_{ab}$ are the electric and magnetic components of the Weyl (conformal curvature) tensor (see below).

\subsection{Long-range gravity}\label{ssL-RG}
%%%%%%%%%%%%%%%%%%%%%%%%%%%%%%%%%%%%%%%%%%%%%
General relativity interprets gravity as the result of spacetime curvature, while the spacetime itself is treated as a 4-dimensional Riemannian manifold. More specifically, the part of the gravitational field which is caused by the local presence of matter is monitored by the Ricci component of the Riemann curvature tensor. Gravity at a distance, on the other hand, is governed by the Weyl part of the Riemann tensor. This splitting of the gravitational field into its local and long-range components is reflected in the decomposition
\begin{equation}
R_{abcd}= C_{abcd}+ \frac{1}{2}\left(g_{ac}R_{bd}+g_{bd}R_{ac}-g_{bc}R_{ad}-g_{ad}R_{bc} \right)- \frac{1}{6}\,R\left(g_{ac}g_{bd}-g_{ad}g_{bc}\right)\,. \label{riemtens}
\end{equation}
In the above, $R_{abcd}$ is the Riemann tensor, $C_{abcd}$ is the Weyl tensor, $R_{ab}=R^c{}_{acb}$ and $R=R^a{}_a$ are the Ricci tensor and the Ricci scalar respectively and $g_{ab}$ the spacetime metric. By construction, $R_{abcd}=R_{cdab}$, $R_{abcd}=R_{[ab][cd]}$, $R_{a[bcd]}=0$ and $R_{ab}=R_{(ab)}$. Also, the Weyl tensor obeys the symmetries of the Riemann tensor and it is traceless as well.

Local gravity is described by the Ricci field and it is monitored by the Einstein field equations. The Weyl tensor, on the other hand, satisfies the Bianchi identities,
\begin{equation}
\nabla^dC_{abcd}= \nabla_{[b}R_{a]c}+ \frac{1}{6}\,g_{c[b}\nabla_{a]}R\,,  \label{bianchi2}
\end{equation}
which could be seen as the field equations of the nonlocal gravitational field, that is of tidal forces and gravitational waves. Within the framework of the 1+3 covariant formalism, the Weyl tensor splits into its electric and magnetic components, defined by
\begin{equation}
E_{ab}= C_{acbd}u^cu^d \hspace{10mm} {\rm and} \hspace{10mm}
H_{ab}= \frac{1}{2}\,\varepsilon_a{}^{cd}C_{cdbe}u^e\,,
\label{WeylEM}
\end{equation}
respectively~\cite{TCM,EMM}. Employing the $E_{ab}$ and $H_{ab}$ tensors, which are both spatial, symmetric and trace-free (i.e.~$E_{ab}u^b=0=H_{ab}u^b$, $E_{ab}=E_{\langle ab\rangle}$ and $H_{ab}=H_{\langle ab\rangle}$), the Weyl tensor decomposes into its irreducible parts according to
\begin{equation}
C_{ab}{}^{cd}= 4\left(u_{[a}u^{[c}+h_{[a}{}^{[c}\right)E_{b]}{}^{d]}+ 2\varepsilon_{abe}u^{[c}H^{d]e}+ 2u_{[a}H_{b]e}\varepsilon^{cde}\,.  \label{weyld}
\end{equation}

Once the $u_a$-field has been introduced, the Bianchi identities split into their temporal and spatial parts. Then, by means of decomposition (\ref{weyld}) the timelike component of the Bianchi identities provides the propagation formulae~\cite{TCM}
\begin{eqnarray}
\dot{E}_{\langle ab\rangle}&=& -\Theta E_{ab}- \frac{1}{2}\,(\rho+p)\sigma_{ab}+ \mbox{curl}H_{ab}- \frac{1}{2}\,\dot{\pi}_{ab}- \frac{1}{6}\,\Theta\pi_{ab}- \frac{1}{2}\,{\rm D}_{\langle a}q_{b\rangle}- A_{\langle a}q_{b\rangle}\nonumber\\
&&+3\sigma_{\langle a}{}^c\left(E_{b\rangle c}- \frac{1}{6}\,\pi_{b\rangle c}\right)+ \varepsilon_{cd\langle a} \left[2A^cH_{b\rangle}{}^d-\omega^c\left(E_{b\rangle}{}^d +\frac{1}{2}\,\pi_{b\rangle}{}^d\right)\right] \label{Eevolutions}
\end{eqnarray}
and
\begin{eqnarray}
\dot{H}_{\langle ab\rangle}&=& -\Theta H_{ab}- \mbox{curl}E_{ab}+ \frac{1}{2}\,\mbox{curl}\pi_{ab}+ 3\sigma_{\langle a}{}^cH_{b\rangle c}- \frac{3}{2}\,\omega_{\langle a}q_{b\rangle}\nonumber\\
&&-\varepsilon_{cd\langle a}\left(2A^cE_{b\rangle}{}^d -\frac{1}{2}\,\sigma^c{}_{b\rangle}q^d +\omega^cH_{b\rangle}{}^d\right)\,.  \label{Hevolution}
\end{eqnarray}
Similarly, the spacelike part of the decomposed Bianchi identities leads to the constraints
\begin{equation}
{\rm D}^bE_{ab}= \frac{1}{3}\,{\rm D}_a\rho- \frac{1}{2}\,{\rm D}^b\pi_{ab}- \frac{1}{3}\,\Theta q_a+ \frac{1}{2}\,\sigma_{ab}q^b- 3H_{ab}\omega^b+ \varepsilon_{abc}\left(\sigma^b{}_dH^{cd} -\frac{3}{2}\,\omega^bq^c\right)  \label{Econstrain}
\end{equation}
and
\begin{equation}
{\rm D}^bH_{ab}= (\rho+p)\omega_a- \frac{1}{2}\,\mbox{curl}q_a+ 3E_{ab}\omega^b- \frac{1}{2}\,\pi_{ab}\omega^b- \varepsilon_{abc}\sigma^b{}_d \left(E^{cd}+\frac{1}{2}\,\pi^{cd}\right)\,. \label{Hconstrain}
\end{equation}
The above govern the action of gravity at a distance and typically obey wave-like solutions.

\section{Friedmannian cosmologies}\label{FLRWCs}
%%%%%%%%%%%%%%%%%%%%%%%%%%%%%%%%%%%%%%%%%%%%%%%%
The observational data, along with the Copernican principle, imply that the universe is uniform (i.e.~homogeneous and isotropic) on very large scales. These symmetries are consistent with the Robertson-Walker geometry, which determines the metric tensor, and with a perfect-fluid energy-momentum tensor for the matter. All these suggest that the FLRW models provide a good description of our universe on sufficiently large scales.

\subsection{The FLRW models}\label{ssFLRWMs}
%%%%%%%%%%%%%%%%%%%%%%%%%%%%%%%%%%%%%%%%%%%%
The uniformity of the FLRW spacetimes demands that any quantity which induces inhomogeneity or anisotropy must vanish identically. This in turn guarantees that only time-dependent scalars survive in the related equations. As a result, the only non-trivial relations emerging from the Einstein field-equations are Raychadhuri's formula
\begin{equation}
\dot{H}= -H^2- \frac{1}{6}\,(\rho+3p)+ \frac{1}{3}\,\Lambda\,, \label{FLRWdotH}
\end{equation}
together with the continuity equation
\begin{equation}
\dot{\rho}= -3H(\rho+p)\,,  \label{FLRWcon}
\end{equation}
with $H=\dot{a}/a$ representing the Hubble parameter and $\Lambda$ the cosmological constant. An additional important formula is the integral of (\ref{FLRWdotH}), namely
\begin{equation}
H^2= \frac{1}{3}\,\rho- \frac{K}{a^2}+ \frac{1}{3}\,\Lambda\,,
\label{FLRWfried}
\end{equation}
where $K=0,\pm1$ is the curvature index of the spatial sections. The above, together with \eqref{FLRWdotH}, comprise the so called Friedmann equations.

\subsection{The Einstein-de Sitter universe}\label{ssE-dSU}
%%%%%%%%%%%%%%%%%%%%%%%%%%%%%%%%%%%%%%%%%%%%%%%%%%%%%%%%%%%
The simplest of the FLRW spacetimes is the so-called Einstein-de Sitter universe, which is spatially flat, has zero cosmological constant and contains pressureless matter (i.e.~`dust'). This model is believed to provide an accurate description of our cosmos from the time of equipartition until the onset of the recent accelerated phase.

Setting $K=0=\Lambda=p$ on to the right-hand side of Eqs.~(\ref{FLRWcon}) and (\ref{FLRWdotH}), the latter lead to the familiar solution
\begin{equation}
\rho= \rho_0\left(\frac{a_0}{a}\right)^3 \hspace{10mm} {\rm and} \hspace{10mm} a= a_0\left(\frac{t}{t_0}\right)^{2/3}\,,
\label{EdSsf}
\end{equation}
with the zero-suffix indicating the beginning of the dust era. Then, it is straightforward to show that $H=2/3t$ and $\rho=4/3t^2$ throughout that period. In what follows, the Einstein-de Sitter model will form our unperturbed (zero-order) background universe.

\section{Cosmological gravitational waves}\label{sCGWs}
%%%%%%%%%%%%%%%%%%%%%%%%%%%%%%%%%%%%%%%%%%%%%%%%%%%%%%%
Typical inflationary models produce gravitational waves, the relative strength of which depends primarily on the energy scale of inflation and on the reheat temperature. These waves cross outside the Hubble horizon during the exponential expansion phase and remain there before re-entering at some later epoch. The earlier the time of exit the later that of re-entry.

\subsection{Isolating the gravitational waves}\label{ssIGWs}
%%%%%%%%%%%%%%%%%%%%%%%%%%%%%%%%%%%%%%%%%%%%%%%%%%%%%%%%%%%%
The long-range gravitational field is monitored by the electric and magnetic parts of the Weyl tensor, the transverse part of which can and has been used to describe cosmological gravitational waves~\cite{H}. When the cosmic medium is a perfect fluid and the background universe is an FLRW model, the aforementioned Weyl components obey the following set of linear relations
\begin{equation}
\dot{E}_{ab}= -3HE_{ab}-{1\over2}\,(\rho+p)\sigma_{ab}+ {\rm curl}H_{ab}\,, \hspace{15mm} \dot{H}_{ab}= -3HH_{ab}- {\rm curl}E_{ab}\,,  \label{ldotEH}
\end{equation}
\begin{equation}
{\rm D}^bE_{ab}= {1\over3}\,{\rm D}_a\rho \hspace{15mm} {\rm and} \hspace{15mm} {\rm D}^bH_{ab}= (\rho+p)\omega_a\,.  \label{lDivEH}
\end{equation}
The first two of the above expressions are propagation formulae, describing the time-evolution of the Weyl field. Equations (\ref{lDivEH}a) and (\ref{lDivEH}b), on the other hand, are constraints satisfied on the observers' 3-D rest-space at all times.

Gravitational waves are pure-tenor perturbations, which means that the related information is encoded in the transverse part of the Weyl field. Therefore, to isolate the gravitational waves, one needs to ensure that the transversality condition is always preserved. Mathematically, this is achieved by imposing the linear 3-D constraints
\begin{equation}
{\rm D}^bE_{ab}= 0 \hspace{15mm} {\rm and} \hspace{15mm} {\rm D}^bH_{ab}= 0\,,  \label{lcon}
\end{equation}
at all times. The above appear to hold as long as ${\rm D}_a\rho=0=\omega_a$ to first approximation, which means to (artificially) switch off the linear effects of density inhomogeneities and vorticity. Nevertheless, to guarantee that transversality is preserved in time, once it is imposed on the initial hypersurface, one needs to demand that ${\rm D}_a\Theta=0$ at the same perturbative level as well.

\subsection{The role of the shear tensor}\label{ssRST}
%%%%%%%%%%%%%%%%%%%%%%%%%%%%%%%%%%%%%%%%%%%%%%%%%%%%%%
Shear perturbations describe shape distortions analogous to those induced by propagating gravitational waves. On a flat FLRW background, the linear shear tensor evolves according to
\begin{equation}
\dot{\sigma}_{ab}= -2H\sigma_{ab}- E_{ab}\,,  \label{ldotsh}
\end{equation}
while it satisfies the associated 3-D constraints
\begin{equation}
{\rm D}^b\sigma_{ab}= {2\over3}\,{\rm D}_a\Theta+ {\rm curl}\omega_a\,, \hspace{15mm} H_{ab}= {\rm curl}\sigma_{ab}+ {\rm D}_{\langle a}\omega_{b\rangle}  \label{lshcon1}
\end{equation}
and the linearised Gauss-Codacci equation \cite{TCM}
\begin{equation}
\mathcal{R}_{\langle ab\rangle}= -H\sigma_{ab}+ E_{ab}\,,  \label{lshcon2}
\end{equation}
where $\mathcal{R}_{\langle ab\rangle}$ is the symmetric and trace-free part of the (perturbed) 3-D Ricci tensor. Following (\ref{lshcon1}a), when ${\rm D}_a\rho=0={\rm D}_a\Theta=\omega_a$, the shear is also transverse (i.e.~${\rm D}^b\sigma_{ab}=0$ to first order) and the same is also true for $\mathcal{R}_{\langle ab\rangle}$ (see constraint (\ref{lshcon2})).

What is more important is that, under these constrains, the linear shear evolution dictates that of the magnetic Weyl tensor (see Eq.~(\ref{lshcon1}b)) and then (by means of (\ref{ldotEH}a) -- see also (\ref{ldotsh})) that of its electric counterpart. Consequently, on an FLRW background, the linear evolution of gravitational-wave perturbations is solely monitored by the shear. The latter obeys the formula
\begin{equation}
\ddot{\sigma}_{ab}= -5H\dot{\sigma}_{ab}- {1\over2}\left(\rho-3p-{2K\over a^2}\right)\sigma_{ab}+ {\rm D}^2\sigma_{ab}\,,  \label{lddotsh1}
\end{equation}
obtained after taking the time derivative of (\ref{ldotsh}) and then using (\ref{ldotEH}a), together with the vorticity-free version of (\ref{lshcon1}b) and the linear relation ${\rm curl}H_{ab}=(3K/a^2)\sigma_{ab}-{\rm D}^2\sigma_{ab}$ (e.g.~see \S~3.6 in~\cite{TCM}). Note that ${\rm D}^2={\rm D}^a{\rm D}_a$ is the 3-D covariant Laplacian operator. The above is a wave-like equation with additional terms due to the expansion, the presence of matter and the nonzero background curvature.

\subsection{Linear gravitational waves}\label{ssLGWs}
%%%%%%%%%%%%%%%%%%%%%%%%%%%%%%%%%%%%%%%%%%%%%%%%%%%%%
After equilibrium we are dealing with non-relativistic matter of zero pressure (i.e.~$p=0$). Therefore, assuming spatial flatness in the background (i.e.~setting $K=0$) and then harmonically decomposing Eq.~(\ref{lddotsh1}), we arrive at
\begin{equation}
\ddot{\sigma}_{(n)}= -5H\dot{\sigma}_{(n)}- {3\over2}\,H^2 \left[1+{2\over3}\left({\lambda_H\over\lambda_n}\right)^2\right] \sigma_{(n)}\,,  \label{lddotsh2}
\end{equation}
since $3H^2=\rho$ to zero order. Here, we have introduced the splitting $\sigma_{ab}=\sum_n\sigma_{(n)} \mathcal{Q}_{ab}^{(n)}$, with ${\rm D}_a\sigma_{(n)}=0$ and $\mathcal{Q}_{ab}^{(n)}$ representing (pure) tensor harmonic functions (i.e.~$\mathcal{Q}_{ab}=\mathcal{Q}_{\langle ab\rangle}$, ${\rm D}^b\mathcal{Q}_{ab}=0=\dot{\mathcal{Q}}_{ab}$ and ${\rm D}^2\mathcal{Q}_{ab}=-(n/a)^2\mathcal{Q}_{ab}$). Also, $\lambda_H=1/H$ and $\lambda_n=a/n$ are the Hubble radius and the physical wavelength of the shear perturbation respectively (with $n>0$ being the comoving wavenumber).

After equipartition we have $a\propto t^{2/3}$ and $H=2/3t$. Then, on superhorizon scales (i.e.~for $\lambda_H/\lambda_n\ll1$), Eq.~(\ref{lddotsh2}) simplifies to\footnote{Alternatively, we could have arrived at Eq.~(\ref{lddotsh3}), and therefore at solution (\ref{lsh}), by setting $p=0=K$ in (\ref{lddotsh1}) and dropping the shear Laplacian from the right-hand side of the same expression.}
\begin{equation}
3t^2\ddot{\sigma}_{(n)}+10t\dot{\sigma}_{(n)}+ 2\sigma_{(n)}= 0  \label{lddotsh3}
\end{equation}
and accepts the power-law solution
\begin{equation}
\sigma_{(n)}= \mathcal{C}_1t^{-1/3}+ \mathcal{C}_2t^{-2}\,.  \label{lsh}
\end{equation}
Therefore, during the dust era, gravitationally induced (i.e.~pure-tensor) shear perturbations decay as $\sigma\propto t^{-1/3}$ on super-Hubble scales (e.g.~see~\cite{TCM}). This linear result will be used later to study the effects of gravitational waves on density perturbations at second order.

\section{Cosmological density perturbations}\label{sCDPs}
%%%%%%%%%%%%%%%%%%%%%%%%%%%%%%%%%%%%%%%%%%%%%%%%%%%%%%%%%
Inhomogeneities in the density of the baryonic component typically start growing after recombination, once matter has decoupled from the background radiation field. Perturbations in the CDM sector, on the other hand, may start condensing (much) earlier. Note that the high isotropy of the Cosmic Microwave Background (CMB) suggests that the amplitude of the the former type of distortions at decoupling should be very small (of the order of $10^{-5}$).

\subsection{Describing density inhomogeneities}\label{ssDDIs}
%%%%%%%%%%%%%%%%%%%%%%%%%%%%%%%%%%%%%%%%%%%%%%%%%%%%%%%%%%%%%
Variations in the density distribution of the matter (baryonic or CDM), as seen between two neighbouring observers, are monitored by the dimensionless gradient~\cite{EB}
\begin{equation}
\Delta_a= {a\over\rho}\,{\rm D}_a\rho\,.  \label{Delta-a}
\end{equation}
The latter carries collective information for all three types of density inhomogeneities, namely scalar, vector and (trace-free) tensor. One can decode this information by taking the comoving gradient of $\Delta_a$ and then introducing the splitting
\begin{equation}
\mathcal{D}_{ab}= a{\rm D}_b\Delta_a= {1\over3}\,\mathcal{D}h_{ab}+ \mathcal{D}_{\langle ab\rangle}+ \mathcal{D}_{[ab]}\,,  \label{delta-ab}
\end{equation}
with $\mathcal{D}=a{\rm D}^a\Delta_a$, $\mathcal{D}_{\langle ab\rangle}=a{\rm D}_{\langle b}\Delta_{a\rangle}$ and $\mathcal{D}_{[ab]}=a{\rm D}_{[b}\Delta_{a]}$ by definition~\cite{EBH}. The scalar $\mathcal{D}$ describes what we commonly refer to as density perturbations, namely overdensities or underdensities in the matter distribution, and closely corresponds to the more familiar density contrast $\delta=\delta\rho/\rho$ (e.g.~see~\cite{TCM,EB}). The symmetric and trace-free tensor $\mathcal{D}_{\langle ab\rangle}$ monitors changes in the shape of the inhomogeneity (under constant volume). Finally, the antisymmetric tensor $\mathcal{D}_{[ab]}$ is related to rotational perturbations (e.g.~vortices) in the matter density.

\subsection{Linear density inhomogeneities}\label{ssLDIs}
%%%%%%%%%%%%%%%%%%%%%%%%%%%%%%%%%%%%%%%%%%%%%%%%%%%%%%%%%
Consider a perturbed spatially flat FLRW spacetime, with zero cosmological constant, containing a single barotropic fluid. To linear order, inhomogeneities in the density distribution of the matter evolve according to the system~\cite{EBH}
\begin{equation}
\dot{\Delta}_a= 3wH\Delta_a- (1+w)Z_a  \label{ldotDelta-a}
\end{equation}
and
\begin{equation}
\dot{Z}_a= -2HZ_a- {1\over2}\,\rho\Delta_a- {3\over2}\,a{\rm D}_ap- a\left[3H^2+{1\over2}\,\rho(1+3w)\right]A_a+ a{\rm D}_a{\rm D}^bA_b\,,  \label{ldotZ-a}
\end{equation}
where the gradient $Z_a=a{\rm D}_a\Theta$ describes spatial variations in the universal volume expansion and $w=p/\rho$ is the barotropic index of the matter~\cite{EBH}. Also, the 4-acceleration satisfies the linear conservation law~\cite{TCM,EMM}
\begin{equation}
\rho(1+w)A_a= -{\rm D}_ap\,.  \label{lmdcl}
\end{equation}

After recombination, and until the onset of the recent accelerated phase, the energy density of the universe is dominated by pressureless dust. Assuming matter with zero pressure (baryonic or CDM), means that the barotropic index and the 4-acceleration vanish as well (i.e.~$w=0=A_a$). Then, the set of (\ref{ldotDelta-a}) and (\ref{ldotZ-a}) reduces to
\begin{equation}
\dot{\Delta}_{(k)}= -Z_{(k)} \hspace{10mm} {\rm and} \hspace{10mm}
\dot{Z}_{(k)}= -2HZ_{(k)}- {1\over2}\,\rho\Delta_{(k)}\,,  \label{ld1}
\end{equation}
after harmonically decomposing the perturbations~\cite{EB}. In particular, after setting $\Delta_a= \Delta_{(k)}\mathcal{Q}_a^{(k)}$ and $Z_a= Z_{(k)}\mathcal{Q}_a^{(k)}$, with ${\rm D}_a\Delta_{(k)}=0={\rm D}_aZ_{(k)}$ and $\dot{\mathcal{Q}}_a^{(k)}=0$, where $\mathcal{Q}_a^{(k)}$ are standard vector harmonic functions. Recalling that $H=2/3t$ and $\rho=4/3t^2$ when pressure-free dust dominates the energy density of the universe, the above system accepts the solution
\begin{equation}
\Delta_{(n)}= C_1t^{2/3}+ C_2t^{-1} \hspace{10mm} {\rm and} \hspace{10mm} Z_{(n)}= C_3t^{-1/3}+ C_4t^{-2}\,.  \label{dsols1}
\end{equation}
Therefore, during the dust era, linear inhomogeneities in the density distribution of the matter grow as $\Delta\propto t^{2/3}$, while those in the Hubble expansion decay as $Z\propto t^{-1/3}$.

Taking the comoving 3-gradient of (\ref{ldotDelta-a}) and (\ref{ldotZ-a}), setting the pressure to zero and using the linear commutation laws $(a{\rm D}_b\Delta_a)^{\cdot}=a{\rm D}_b\dot{\Delta}_a$ and $(a{\rm D}_bZ_a)^{\cdot}=a{\rm D}_b\dot{Z}_a$, we arrive at
\begin{equation}
\dot{\mathcal{D}}_{ab}= -\mathcal{Z}_{ab} \hspace{10mm} {\rm and} \hspace{10mm}
\dot{\mathcal{Z}}_{ab}= -2H\mathcal{Z}_{ab}- {1\over2}\,\rho\mathcal{D}_{ab}\,,  \label{ld2}
\end{equation}
with $\mathcal{Z}_{ab}=a{\rm D}_bZ_a$.\footnote{In analogy with (\ref{delta-ab}), the variable $\mathcal{Z}_{ab}$ splits into its trace, its antisymmetric and its symmetric and trace-free component according to $\mathcal{Z}_{ab}= (\mathcal{Z}/3)h_{ab}+\mathcal{Z}_{\langle ab\rangle}+ \mathcal{Z}_{[ab]}$, where $\mathcal{Z}=a{\rm D}^aZ_a$, $\mathcal{Z}_{\langle ab\rangle}=a{\rm D}_{\langle b}Z_{a\rangle}$ and $\mathcal{Z}_{[ab]}=a{\rm D}_{[b}Z_{a]}$.} Introducing the harmonic splitting $\mathcal{D}_{ab}=\mathcal{D}_{(k)}\mathcal{Q}_{ab}^{(k)}$ and $\mathcal{Z}_{ab}=\mathcal{Z}_{(k)}\mathcal{Q}_{ab}^{(k)}$, where ${\rm D}_a\mathcal{D}_{(k)}=0={\rm D}_a\mathcal{Z}_{(k)}$ and $\dot{\mathcal{Q}}_a^{(k)}=0$, with $\mathcal{Q}_a^{(k)}$ being standard tensor harmonic functions, the above systems solves to give
\begin{equation}
\mathcal{D}_{(n)}= \tilde{C}_1t^{2/3}+ \tilde{C}_2t^{-1} \hspace{10mm} {\rm and} \hspace{10mm} \mathcal{Z}_{(n)}= \tilde{C}_3t^{-1/3}+ \tilde{C}_4t^{-2}\,.  \label{dsols2}
\end{equation}

According to solutions (\ref{dsols1}a) and (\ref{dsols2}a), during the dust era, all types of linear density inhomogeneities, namely overdensities/underdensities, shape distortions and vortices, obey the same evolution law (see also~\cite{MTM}). Moreover, following solutions (\ref{dsols1}b) and (\ref{dsols2}b), the same is also true for the linear inhomogeneities in the expansion.

\section{Allowing for gravitational-wave %%%%%%%%%%%%%%%%%%%%%%%%%%%%%%%%%%%%%%%%
effects}\label{sAG-WEs}
%%%%%%%%%%%%%%%%%%%%%%%
At the linear level, there is no coupling between gravitational-wave distortions and density perturbations. In order to investigate the effects of the Weyl field, one needs to go to a higher perturbative level and in particular to include second-order terms into the equations.

\subsection{Nonlinear density perturbations}\label{ssN-LDPs}
%%%%%%%%%%%%%%%%%%%%%%%%%%%%%%%%%%%%%%%%%%%%%%%%%%%%%%%%%%%%
Suppose that the cosmic medium is a single barotropic (perfect) fluid. Then, for zero rotation and in the absence of a cosmological constant, inhomogeneities in the density distribution of the matter are monitored by the nonlinear propagation formulas \cite{TCM}
\begin{equation}
\dot{\Delta}_{\langle a\rangle}= w\Theta\Delta_a- (1+w)Z_a- \sigma_{ab}\Delta^b  \label{nldotDelta-a}
\end{equation}
and
\begin{eqnarray}
\dot{Z}_{\langle a\rangle}&=& -{2\over3}\,\Theta Z_a- {1\over2}\,\rho\Delta_a- {3\over2}\,a{\rm D}_ap- a\left[{1\over3}\,\Theta^2+{1\over2}\,(\rho+3p)\right]A_a+ a{\rm D}_a{\rm D}^bA_b \nonumber\\ &&-\sigma_{ab}Z^b- 2a{\rm D}_a\sigma^2+ 2aA^b{\rm D}_aA_b+ a({\rm D}^bA_b)A_a- a\left(2\sigma^2-A^bA_b\right)A_a\,,  \label{nldotZ-a}
\end{eqnarray}
where $\dot{\Delta}_{\langle a\rangle}=h_a{}^b\dot{\Delta}_a$ and $\dot{Z}_{\langle a\rangle}=h_a{}^b\dot{Z}_a$ by definition. However, when dealing with a single perfect (barotropic) fluid, the nonlinear momentum-density conservation law (see~Eq.~(\ref{lmdcl}) in \S~\ref{ssLDIs}) ensures that the 4-acceleration vanishes for zero pressure. In that case it is straightforward to show that $\dot{\Delta}_{\langle a\rangle}=\dot{\Delta}_a$ and $\dot{Z}_{\langle a\rangle}=\dot{Z}_a$. Therefore, after matter-radiation equality, the system (\ref{nldotDelta-a}) and (\ref{nldotZ-a}) simplifies to
\begin{equation}
\dot{\Delta}_a= -Z_a- \sigma_{ab}\Delta^b  \label{nldotDelta-a2}
\end{equation}
and
\begin{equation}
\dot{Z}_a= -{2\over3}\,\Theta Z_a- {1\over2}\,\rho\Delta_a- \sigma_{ab}Z^b- 2a{\rm D}_a\sigma^2\,,  \label{nldotZ-a2}
\end{equation}
respectively. The above set, together with the associated propagation formula of the shear (see \S~\ref{ssKs} earlier), monitors the nonlinear evolution of inhomogeneities in the density of a universe dominated by pressureless matter (baryonic or/and CDM).

\subsection{The second-order interaction}\label{ssS-OI}
%%%%%%%%%%%%%%%%%%%%%%%%%%%%%%%%%%%%%%%%%%%%%%%%%%%%%%%
In what follows we will focus exclusively on the role of gravitational-wave distortions, as these propagate through the transverse component of the shear, on density perturbations. More specifically, we will consider the effects of these distortions at the second perturbative level. We will therefore need to introduce a background model, which in our case will coincide with the Einstein-de Sitter universe (see \S~\ref{ssE-dSU} earlier). Within this approximation scheme the nonlinear expressions (\ref{nldotDelta-a2}) and (\ref{nldotZ-a2}) reduce to
\begin{equation}
\dot{\Delta}_a+ Z_a= -\tilde{\sigma}_{ab}\tilde{\Delta}^b  \label{2dotDelta-a}
\end{equation}
and
\begin{equation}
\dot{Z}_a+ 2HZ_a+ {1\over2}\,\bar{\rho}\Delta_a= -\tilde{\sigma}_{ab}\tilde{Z}^b- 2a{\rm D}_a\tilde{\sigma}^2\,,  \label{2dotZ-a}
\end{equation}
respectively. Note that the variables $\Delta_a$ and $Z_a$ (and their temporal derivatives), seen on the left hand sides of the above are treated as second order perturbations, which makes $H=\bar{\Theta}/3$ the background Hubble parameter and $\bar{\rho}$ the associated matter density. Consequently, to maintain the 2nd perturbative order of both (\ref{2dotDelta-a}) and (\ref{2dotZ-a}), the variables $\tilde{\sigma}_{ab}$, $\tilde{\Delta}_a$, $\tilde{Z}_a$ and $\tilde{\sigma}$ seen on the right-hand sides will be treated a linear perturbations. Hence, here onwards, ``overbars'' will always indicate zero-order variables and `tildas'' first-order ones. In addition, assuming that the shear is entirely due to gravitational-wave distortions, we may only account for its divergence-free component (i.e.~set ${\rm D}^b\tilde{\sigma}_{ab}=0$ to first order). We finally note that, in accord with Eq.~(\ref{2dotDelta-a}), gravitational waves do not directly induce density inhomogeneities to second order. Nevertheless, the last term of (\ref{2dotZ-a}) suggests that the shear gradients can in principle trigger distortions in the volume expansion, which then can lead to density perturbations (see Eq.~\ref{2dotDelta-a})).

The rest of this work looks into the second-order effects of gravitational waves on pre-existing (scalar) density perturbations, namely on overdensities/underdensities in the matter distribution (baryonic or/and CDM). Thus, following \S~\ref{ssDDIs} (see also footnote~4 in \S~\ref{ssLDIs}), our next step is to take the comoving 3-divergence of Eqs.~(\ref{2dotDelta-a}) and (\ref{2dotZ-a}). On doing so, and after keeping up to second-order terms, we arrive at\footnote{The second-order expressions (\ref{2dotDelta}) and (\ref{2dotZ}) have been derived after introducing the Einstein-de Sitter background, obtaining the 2nd-order limit of Eqs.~(\ref{nldotDelta-a2}), (\ref{nldotZ-a2}) and then taking the comoving 3-divergences of the latter. Strictly speaking, however, one should take the 3-divergences of the nonlinear expressions (\ref{nldotDelta-a2}), (\ref{nldotZ-a2}) first, and then introduce the Einstein-de Sitter background that will eventually lead to the 2nd-order propagation equations of $\mathcal{D}$ and $\mathcal{Z}$. Nevertheless, as well will show in Appendix~\ref{sA2}, the aforementioned difference in the derivation does not alter the final outcome of the analysis. For this reason, here, we have chosen to follow the simplest approach.}
\begin{equation}
\dot{\mathcal{D}}+ \mathcal{Z}= -2\tilde{\sigma}_{ab} \tilde{\mathcal{D}}^{\langle ab\rangle}  \label{2dotDelta}
\end{equation}
and
\begin{equation}
\dot{\mathcal{Z}}+ 2H\mathcal{Z}+ {1\over2}\,\bar{\rho}\,\mathcal{D}= -2\tilde{\sigma}_{ab}\tilde{\mathcal{Z}}^{\langle ab\rangle}- 2a^2{\rm D}^2\tilde{\sigma}^2\,.  \label{2dotZ}
\end{equation}
Note that in deriving the above, we have accounted for the divergence-free nature of the gravitationally induced shear (i.e.~the fact that ${\rm D}^b\tilde{\sigma}_{ab}=0$ at the linear perturbative level -- see~\S~\ref{ssRST} earlier). We have also employed the auxiliary second-order commutation laws $a{\rm D}^a\dot{\Delta}_a=\dot{\mathcal{D}}+ \tilde{\sigma}_{ab}\tilde{\mathcal{D}}^{\langle ab\rangle}$ and $a{\rm D}^a\dot{Z}_a=\dot{\mathcal{Z}}+ \tilde{\sigma}_{ab}\tilde{\mathcal{Z}}^{\langle ab\rangle}$ (see Appendix~A for details on their derivation).

Focusing on supperhorizon-sized perturbations, with $\lambda/\lambda_H\ll1$ at all times, we can safely ignore the shear Laplacian on the right-hand side of Eq.~(\ref{2dotZ}). Then, taking the time derivative of (\ref{2dotDelta}) and using (\ref{2dotZ}), together with the linear evolution laws (\ref{ldotsh}), (\ref{ld2}a) and the first-order constraint (\ref{lshcon2}), gives
\begin{equation}
\ddot{\mathcal{D}}+ 2H\dot{\mathcal{D}}- {1\over2}\,\bar{\rho}\,\mathcal{D}= 2H\tilde{\sigma}_{ab}\tilde{\mathcal{D}}^{\langle ab\rangle}+ 4\tilde{\sigma}_{ab}\tilde{\mathcal{Z}}^{\langle ab\rangle}+ 2\tilde{\mathcal{R}}_{\langle ab\rangle}\tilde{\mathcal{D}}^{\langle ab\rangle}\,.  \label{2ddotcD1}
\end{equation}
This differential equation monitors the evolution of large-scale scalar density perturbations (i.e.~overdensities/underdensities), driven by gravitational-wave distortions, at the second perturbative level. To solve the above, we recall that $\tilde{\sigma}_{ab}\propto t^{-1/3}$, $\tilde{\mathcal{D}}_{\langle ab\rangle}\propto t^{2/3}$, $\tilde{\mathcal{Z}}_{\langle ab\rangle}\propto t^{-1/3}$ and $\tilde{\mathcal{R}}_{\langle ab\rangle}\propto t^{-4/3}$ to first order (see \S~\ref{ssLGWs} and \S~\ref{ssLDIs} earlier). Then, without any loss of generality, the second-order driving terms on the right-hand side of Eq.~(\ref{2ddotcD1}) can be written in the form
\begin{equation}
\tilde{\sigma}_{ab}\tilde{\mathcal{D}}^{\langle ab\rangle}= \tilde{\sigma}_0\tilde{\mathcal{D}}_0\left({t\over t_0}\right)^{1/3}\,, \hspace{5mm} \tilde{\sigma}_{ab}\tilde{\mathcal{Z}}^{\langle ab\rangle}= \tilde{\sigma}_0\tilde{\mathcal{Z}}_0\left({t_0\over t}\right)^{2/3}  \label{2dts1}
\end{equation}
and
\begin{equation}
\tilde{\mathcal{R}}_{\langle ab\rangle} \tilde{\mathcal{D}}^{\langle ab\rangle}= \tilde{\mathcal{R}}_0\tilde{\mathcal{D}}_0\left({t_0\over t}\right)^{2/3}\,,  \label{2dts2}
\end{equation}
where the zero suffix denotes the time of recombination.\footnote{Using the linear evolution laws of the shear and the density perturbation, the scalar $\tilde{\sigma}_{ab}\tilde{\mathcal{D}}^{\langle ab\rangle}$ reads $\tilde{\sigma}_{ab}\tilde{\mathcal{D}}^{\langle ab\rangle}= (\tilde{\sigma}_{11}^0 \tilde{\mathcal{D}}_0^{\langle 11\rangle}+ \tilde{\sigma}_{22}^0\tilde{\mathcal{D}}_0^{\langle 22\rangle}+ \tilde{\sigma}_{33}^0\tilde{\mathcal{D}}_0^{\langle 33\rangle}) (t/t_0)^{1/3}$. Then, the sum $\tilde{\sigma}_{ab} \tilde{\mathcal{D}}^{\langle ab\rangle}$ can always be written in the form (\ref{2dts1}a). Note that we have assumed (for simplicity) that the $3\times3$ matrices $\tilde{\sigma}_{ab}$ and $\tilde{\mathcal{D}}^{\langle ab\rangle}$ are diagonal. Also, the quantities $\tilde{\sigma}_{11}^0$, $\tilde{\mathcal{D}}_0^{\langle 11\rangle}$, etc, have been evaluated at $t=t_0=t_{RC}$ and they are generally functions of space. Similarly, one can show that the sums $\tilde{\sigma}_{ab}\tilde{\mathcal{Z}}^{\langle ab\rangle}$ and $\tilde{\mathcal{R}}_{\langle ab\rangle}\tilde{\mathcal{D}}^{\langle ab\rangle}$ can be written in the form (\ref{2dts1}b) and (\ref{2dts2}) respectively.} The latter marks the decoupling between ordinary matter and radiation and the onset of (baryonic) structure formation. Finally, recalling that $H=2/3t$ and $\bar{\rho}=4/3t^2$ after equipartition, expressions (\ref{2ddotcD1})-(\ref{2dts2}) combine to
\begin{equation}
3t^2\ddot{\mathcal{D}}+ 4t\,\dot{\mathcal{D}}- 2\mathcal{D}= 2\alpha_0\left({t\over t_0}\right)^{4/3}\,,  \label{2ddotcD2}
\end{equation}
with $\alpha_0$ being a dimensionless parameter. This has been evaluated at recombination and carries the (2nd-order) driving effect of the gravitational waves on $\mathcal{D}$. Also, $\alpha_0$ can generally vary in space and it is given by
\begin{equation}
\alpha_0= 2\tilde{\sigma}_0\tilde{\mathcal{D}}_0t_0+ 6\tilde{\sigma_0}\tilde{\mathcal{Z}}_0t_0^2+ 3\tilde{\mathcal{R}}_0\tilde{\mathcal{D}}_0t_0^2\,.  \label{2alpha01}
\end{equation}
The set of (\ref{2ddotcD2}) and (\ref{2alpha01}) governs the evolution of (scalar) density perturbations in the presence of cosmological gravitational waves, on superhorizon scales and during the post-recombination era, to second perturbative order. Equation (\ref{2ddotcD2}) can be solved analytically, but before we proceed to its solution we should first consider the so-called ``gauge issue''.

\subsection{The gauge issue}\label{ssGI}
%%%%%%%%%%%%%%%%%%%%%%%%%%%%%%%%%%%%%%%%
Cosmological perturbation theory is generally susceptible to gauge-related problems~\cite{LK}. These stem from the fact that, technically speaking, when studying cosmological perturbations, we deal with two spacetimes. The first is the (fictitious) background spacetime ($\bar{\mathcal{W}}$), which usually coincides with one of the FLRW models. The second is the perturbed spacetime ($\mathcal{W}$) that is thought to provide a more realistic description of the actual universe. To proceed, one needs to introduce an one-to-one mapping (i.e.~a ``gauge'') between the aforementioned two spacetimes. Changing the aforementioned correspondence, while keeping the background spacetime fixed, is known as a gauge transformation. The latter differs from an ordinary coordinate transformation because it changes the event in the background spacetime that corresponds to a given event in its physical counterpart. As a result, the solutions of the perturbed differential equations may be gauge-dependent and they may contain spurious gauge-modes of no real physical substance (e.g.~see~\cite{B}). The 1+3 covariant approach to cosmological perturbations bypasses the gauge problem by using gauge-invariant variables (see~\cite{EB} for details and also for an illuminating discussion of the gauge issue in cosmology). At the linear level, this is achieved by appealing to the Stewart~\&~Walker lemma~\cite{SW}. According to the latter, a first-order variable is gauge-invariant when it vanishes in the background, or when it is a constant scalar there. Following~\cite{BMMS}, when the aforementioned requirements also hold at the linear level, the variable in question is gauge-invariant at second order as well.

The variables used in our linear analysis satisfy the criteria for gauge-invariance, given the uniformity of our FLRW background. This is no longer true at second order, however, since the density contrast ($\mathcal{D}$) has both temporal and spatial dependence at the linear level. To bypass this mathematical problem we will from now on only consider the homogeneous component of the linear density perturbations (i.e. assume that $\mathcal{D}=\mathcal{D}(t)$ to first order).\footnote{Only the 3-divergence ($\mathcal{D}=a{\rm D}^a\Delta_a$) of $\Delta_a$ is assumed to be spatially homogeneous at the linear level. To second order, the scalar $\mathcal{D}$ is allowed to vary both in time and in space. Also, the linear spatial gradient $\Delta_a=(a/\rho){\rm D}_a\rho$ has both temporal and spatial dependence. In cosmology, the best known example of a spatially homogeneous 3-divergence obtained from an inhomogeneous vector field is perhaps the Hubble parameter ($H$, with $3H={\rm D}^au_a$) of an FLRW spacetime.} Then, since $\mathcal{D}\propto t^{2/3}$ at the linear level, it follows that the quantity
\begin{equation}
\mathfrak{D}= t^{-2/3}\mathcal{D}\,,  \label{cDelta}
\end{equation}
will be constant to first order. Therefore, according to~\cite{BMMS}, the above defined scalar is gauge-invariant at the second perturbative level (where it can have both temporal and spatial dependence). Moreover, starting from Eq.~(\ref{2ddotcD2}), it is straightforward to show that the new variable satisfies the second-order differential equation
\begin{equation}
3t^2\ddot{\mathfrak{D}}+ 8t\dot{\mathfrak{D}}= 2\beta_0\left({t\over t_0}\right)^{2/3}\,,  \label{2ddotcD}
\end{equation}
where $\beta_0=\alpha_0/t_0^{2/3}$ is generally a function of space and $\alpha_0$ has been defined in (\ref{2alpha01}). The above accepts the power-law solution, which after evaluating the integration constants reads
\begin{eqnarray}
\mathfrak{D}&=& -{2\over5}\left({\dot{\mathfrak{D}}_0\over H_0} -{3\over7}\,\beta_0\right)\left({t_0\over t}\right)^{5/3} \nonumber\\ &&+ \mathfrak{D}_0+ {2\over5}\,{\dot{\mathfrak{D}}_0\over H_0}- {3\over5}\,\beta_0 \nonumber\\ &&+{3\over7}\,\beta_0\left({t\over t_0}\right)^{2/3}\,,  \label{2ndcD1}
\end{eqnarray}
with $\mathfrak{D}_0$, $\dot{\mathfrak{D}}_0$ and $\beta_0$ being functions of space in general (evaluated at $t=t_0=2/3H_0$).\footnote{Solution (\ref{2ndcD1}) can be also obtained by introducing the variable $\mathfrak{Z}=t^{1/3}\mathcal{Z}$, which is constant at the linear level and therefore gauge-invariant to second order (when $\mathcal{Z}=\mathcal{Z}(t)$ to first-order). Then, on large scales, $3t\dot{\mathfrak{D}}+2\mathfrak{D}+ 3\mathfrak{Z}=-6t^{1/3}\tilde{\sigma}_{ab} \tilde{\mathcal{D}}^{\langle ab\rangle}$ and $3t\dot{\mathfrak{Z}}+3\mathfrak{Z}+ 2\mathfrak{D}=-6t^{4/3}\tilde{\sigma}_{ab} \tilde{\mathcal{Z}}^{\langle ab\rangle}$, which solves analytically to give (\ref{2ndcD1}).} Recasting this solution, which contains one decaying, one constant and one growing mode, in terms of $\mathcal{D}=t^{2/3}\mathfrak{D}$ (see definition (\ref{cDelta})) leads to
\begin{eqnarray}
\mathcal{D}&=& {2\over5}\left(\mathcal{D}_0 -{\dot{\mathcal{D}}_0\over H_0}+{3\over7}\,\alpha_0\right) \left({t_0\over t}\right) \nonumber\\ &&+{3\over5}\left(\mathcal{D}_0+{2\over3}\,{\dot{\mathcal{D}}_0\over H_0}- \alpha_0\right)\left({t\over t_0}\right)^{2/3} \nonumber\\ &&+{3\over7}\,\alpha_0\left({t\over t_0}\right)^{4/3}\,,  \label{2ndcD2}
\end{eqnarray}
guaranteeing that $\mathcal{D}(t=t_0)=\mathcal{D}_0$. According to the above, in the presence of gravitational-wave distortions, (scalar) density perturbations grow ad $\mathcal{D}\propto t^{4/3}\propto a^{2}$ to second order (see \S~\ref{ssIs} next for further discussion on this result). Strictly mathematically speaking, however, solutions (\ref{2ndcD1}) and (\ref{2ndcD2}) are gauge-invariant only for the spatially homogeneous component of the linear density perturbations. Having said that, the gauge-dependence of a variable does not necessarily guarantee the presence of unphysical spurious modes in the associated solution. It is therefore likely that our 2nd-order results cover the whole range of density perturbations and not only those with uniform linear component. We also note that, from the physical point of view, setting $\mathcal{D}=\mathcal{D}(t)$ to first-order means that the scalar $\mathcal{D}$ monitors the average value of the linear density contrast inside the associated overdensity/underdensity; a fairly common practice in nonlinear perturbation studies (e.g.~see \S~8 in~\cite{P}).

\subsection{Implications}\label{ssIs}
%%%%%%%%%%%%%%%%%%%%%%%%%%%%%%%%%%%%%
Ignoring the 2nd-order interaction terms (i.e.~setting $\beta_0=0$) on the right-hand side of (\ref{2ndcD2}) reproduces the standard linear solution, where the dominant mode grows as $\mathcal{D}\propto t^{2/3}\propto a$. When the effects of gravitational radiation are included, however, the second-order solution contains a faster growing mode (with $\mathcal{D}\propto t^{4/3}\propto a^2$). Therefore, qualitatively speaking, the interaction between cosmological gravitational waves and density perturbations (baryonic or/and CDM) increases the growth-rate of the latter. Quantitatively speaking, however, the coefficients of solution (\ref{2ndcD2}) indicate that this new (gravitationally induced) mode will dictate the evolution of density perturbations if $\alpha_0\gtrsim\mathcal{D}_0$, or if $\alpha_0\gtrsim\dot{\mathcal{D}}_0/H_0$, depending which one of $\mathcal{D}_0$ and $\dot{\mathcal{D}}_0/H_0$ is larger. For simplicity, but without any real loss of generality, we may assume that $\mathcal{D}_0\sim\dot{\mathcal{D}}_0/H_0$ and set $\mathcal{D}_0\sim\tilde{\mathcal{D}}_0$ as the initial conditions at the start of the 2nd-order interaction between gravitational waves and density perturbations. Then, the new fast-growing mode on the right-hand side of (\ref{2ndcD2}) will dominate if $\alpha_0\gtrsim\tilde{\mathcal{D}}_0$. However, recalling that $H=2/3t$ after matter-radiation equality, definition (\ref{2alpha01}) recasts into
\begin{equation}
\alpha_0= {4\over3}\,{\tilde{\sigma}_0\over H_0}\,\tilde{\mathcal{D}}_0+ {8\over3}\,{\tilde{\sigma}_0\over H_0}\,\tilde{{\mathcal{Z}}_0\over H_0}+ {4\over3}\,{\tilde{\mathcal{R}}_0\over H_0^2}\, \tilde{\mathcal{D}}_0\,.  \label{alpha02}
\end{equation}
Consequently, since $\tilde{\mathcal{D}}_0$, $\tilde{\sigma}_0/H_0$, $\tilde{\mathcal{Z}}_0/H_0$ and $\tilde{\mathcal{R}}_0/H_0^2$ are all much smaller than unity, we deduce that $\alpha_0\ll \tilde{\mathcal{D}}_0$. In practice, this ensures that the gravitationally induced mode of (\ref{2ndcD2}) can only dominate at very late times. This is not surprising, in view of the (expected) extreme weakness of cosmological gravitational waves (i.e.~the fact that $\tilde{\sigma}_0/H_0\ll1$).\footnote{Since the shear tensor is directly related to the energy density of gravitational-wave perturbations (e.g.~see~\cite{Go}) and the Hubble parameter to the energy density of the dominant background matter, the dimensionless ratio $\tilde{\sigma}/H$ also measures the relative strength between the Weyl and Ricci components of the gravitational field.} More specifically, comparing the two growing modes of solution (\ref{2ndcD2}), shows that Weyl-curvature distortions will start dictating the evolution of large-scale overdensities/underdensities at $t=t_{GW}$, when
\begin{equation}
\alpha_0\left({t_{GW}\over t_0}\right)^{2/3}\gtrsim \tilde{\mathcal{D}}_0\,.  \label{tGW1}
\end{equation}
Then, given that $\alpha_0\sim(\tilde{\sigma}_o/H_0) \tilde{\mathcal{D}}_0$, we estimate that
\begin{equation}
t_{GW}\sim \left({H_0\over\sigma_0}\right)^{3/2}t_0\gg t_0\,,  \label{tGW2}
\end{equation}
where $t_0=t_{RC}\sim10^5$~yrs is the age of the universe at recombination. To obtain a rough estimate of $t_{GW}$, recall that the high isotropy of the CMB demands that $\sigma_0/H_0\lesssim10^{-5}$ at last scattering. Then, Eq.~(\ref{tGW2}) leads to $t_{GW}\gtrsim10^{15/2}t_{RC}$, which translates into $t_{GW}\gtrsim10^{5/2}t_*$, with $t_*\sim10^{10}$~yrs representing the current age of the universe.

Overall, our 2nd-order analysis suggests that the Weyl (rather than the Ricci) component of the gravitational field may become the key factor driving the growth of density perturbations (baryonic or/and CDM) at very late times. Intuitively, one could explain this result by recalling that after equipartition the matter density, which determines the local (i.e.~the Ricci) part of the gravitational field, drops as $\rho\propto a^{-3}$. The divergence-free component of the shear, on the other hand, determines the far (i.e.~the Weyl) field and decays slower, as $\sigma\propto a^{-1/2}$. This means that, although the gravitationally-induced mode on the right-hand side of solution (\ref{2ndcD2}) may be still too weak to have a measurable effect on density perturbations crossing inside the horizon at the present time, it could eventually dominate in the far future.

The results reported here, which have been obtained by means of covariant techniques, are largely in agreement with those obtained in~\cite{G} using Bardeen's formalism, save for the fact that our analysis is confined to the post-equilibrium universe only and to super-Hubble scales. More specifically, it was found in~\cite{G} that long-wavelength relic gravitons contribute a weak growing mode to scalar curvature perturbations, which depends on the ratio between the graviton energy density and that of the dominant matter component. The latter is directly related to the dimensionless ratio $\tilde{\sigma}/H$ that conveys the gravitational-wave effects on density fluctuations here (see footnote~9). Note that, although curvature distortions and density perturbations are not the same, they are related.\footnote{Following~\cite{EMM,MW}, let us consider the (linear) relation $\zeta=-\mathcal{R}-(\mathcal{H}\rho/\rho^{\prime})\delta$, where $\mathcal{H}=aH$ by definition and the prime indicates differentiation with respect to the conformal time. The scalars $\mathcal{R}$ and $\delta$ describe curvature and density perturbations in the comoving-orthogonal gauge. Then, $\zeta$ describes curvature perturbations in the uniform-density gauge and it is directly related to the density perturbations in the spatially-flat gauge. During the dust era $\mathcal{H}\rho/\rho^{\prime}\propto a^{-1}$. Hence, when the curvature distortion scales as $\mathcal{R}\propto a$ and the associated power spectrum as $\mathcal{P}_{\mathcal{R}}\propto a^2$ (as reported in~\cite{G}), the density fluctuation is expected to scale $\delta\propto a^2$ (as found here). For further discussion and details on the so-called Bardeen variables the reader is referred to \S~10.2 in~\cite{EMM} and to \S~7 in~\cite{MW}.} Furthermore, the gravitationally-induced mode reported in~\cite{G} was attributed to the slow decay of the graviton energy density, relative to that of the matter, during both the radiation and the dust eras. The same reasoning was also adopted here (see previous paragraph).

\section{Discussion}\label{sD}
%%%%%%%%%%%%%%%%%%%%%%%%%%%%%%
Structure formation scenarios have a long history in the literature, essentially starting with the pioneering work of Jeans on gravitational instability in the beginning of the last century. Cosmological perturbation theory is the cornerstone upon which all structure formation studies are based. So far, most of the analytical work is confined to the linear regime, The nonlinear approaches are mainly numerical, unless extra simplifying conditions are imposed, with the spherical collapse model and the Zeldovich approximation being the best known examples. Most of the nonlinear studies consider the evolution of structure, once the proto-galactic cloud has ``decoupled'' from the background Hubble expansion, ``turned around'' and started to collapse. Nevertheless, nonlinearities could in principle also affect the early stages of structure formation. On substantially large (superhorizon) scales, all local (causal) physical processes are entirely unimportant. There, the perturbations are primarily affected by the background expansion and, to a lesser degree, by distortions in the curvature of the spacetime. Weyl-curvature distortions, such as those triggered by cosmological gravitational waves, interact with inhomogeneities in the matter density, thought only at second order. Here, we have considered the aforementioned interaction at the second perturbative level and on super-Hubble scales. We did so by assuming an Einstein-de Sitter background, neglecting cosmic rotation and taking into account only the divergence-free (i.e.~the pure-tensor) component of the linear shear perturbations. Our results appear to be in close agreement with those recently reported in~\cite{G}, although the two approaches are quite different formalistically.

Studies of cosmological perturbations are hampered by the so-called gauge issue, which could in principle lead to spurious (gauge-dependent) solutions. To avoid the problem, one could choose the best-fit gauge for the study, or employ a gauge-independent analysis. The latter should involve the use of physically unambiguous gauge-invariant variables, which are however difficult to construct, especially at perturbative levels higher than the linear. In this study, we utilised the results of the linear theory to construct a simple variable that is directly related to (scalar) density perturbations, namely to overdensities/underdensities, and which is also gauge-invariant to second order. Mathematically speaking, we have achieved this by focusing on the uniform (i.e.~the spatially homogeneous) component of the linear density perturbation. Note that it is the only spatial divergence ($\tilde{\mathcal{D}}=a{\rm D}^a\Delta_a$) that is uniform to first order and not the density gradient ($\Delta_a=(a/\rho){\rm D}_a\rho$) itself. Physically, this is like saying that our linear density contrast has been averaged over the overdensity/underdensity, something fairly common in nonlinear perturbation studies. Although strictly theoretically our assumption confines the gauge-invariance of our analysis to a specific subset of density perturbations, namely those with uniform linear density contrast, it is likely that in practice our results cover the full range of these distortions. Alternative gauge-invariant quantities that could rigorously address this issue may be possible to construct, though it is also likely that the physical gain will not compensate for the increased mathematical complexity.

A well known result of the linear study is that, after equipartition, perturbations in the density of the matter (baryonic or CDM) grow in tune with the dimensions of the universe. In other words, $\mathcal{D}\propto t^{2/3}\propto a$, where $a$ is the cosmological scale factor. Our second-order analysis argues that, when the effect of gravitational-wave distortions is also accounted for, the growth rate increases to $\mathcal{D}\propto t^{4/3}\propto a^2$ on superhorizon scales. One might be therefore led to conclude that Weyl-curvature distortions might have substantially accelerated the growth of large-scale density perturbations and thus assisted the formation of structure in the universe. Nevertheless, a closer look reveals that this is very unlikely to have happened because of the (anticipated) extreme weakness of cosmological gravitational waves. The latter essentially guarantees that the aforementioned faster-growing mode can have a measurable effect on the evolution of density perturbations only in the far future. All these might change, however, if an analogous effect was to be observed on subhorizon scales as well. In any case, the possibility that the Weyl, rather than the Ricci, component of the gravitational field could dictate the evolution of density inhomogeneities at late times remains. Intuitively, one could explain this by recalling that the local (the Ricci) field depletes faster than its long-rage (Weyl) counterpart. Then, although Weyl-curvature distortions might play an entirely unimportant role for long, given enough time, they could eventually dominate.

\appendix

\section{Nonlinear commutation laws}\label{sA1}
%%%%%%%%%%%%%%%%%%%%%%%%%%%%%%%%%%%%%%%%%%%%%%%
By definition $a{\rm D}^a\dot{\Delta}_a= ah^{ab}\nabla_b\dot{\Delta}_a$ and $\dot{\Delta}_a=u^b\nabla_b\Delta_a$. Combining these two expressions, employing the Ricci identities and using decomposition (\ref{vdecomposition}) gives
\begin{eqnarray}
a{\rm D}^a\dot{\Delta}_a&=& ah^{ab}\nabla_bu^c\nabla_c\Delta_a+ ah^{ab}u^c\nabla_b\nabla_c\Delta_a\nonumber\\ &=&{1\over3}\,\Theta\Delta+ \left(\sigma^{ab}-\omega^{ab}\right)\Delta_{ab}+ ah^{ab}\left(\nabla_b\Delta_a\right)^{\cdot}+ ah^{ab}u^cR_{bcad}\Delta^d\,,  \label{A1}
\end{eqnarray}
with $R_{abcd}$ representing the Riemann tensor. Using the symmetries of the curvature tensor (see \S~\ref{ssL-RG}), the above leads to the nonlinear commutation law
\begin{eqnarray}
a{\rm D}^a\dot{\Delta}_a&=& \sigma^{ab}\mathcal{D}_{\langle ab\rangle}- \omega^{ab}\mathcal{D}_{[ab]}+ \dot{\mathcal{D}}- aA^a\dot{\Delta}_a- au^aA^b\nabla_b\Delta_a+ au^aR_{ab}\Delta^b \nonumber\\ &=& \dot{\mathcal{D}}+ \sigma^{ab}\mathcal{D}_{\langle ab\rangle}- \omega^{ab}\mathcal{D}_{[ab]}+ {1\over3}\,a\Theta A^a\Delta_a- aA^a\dot{\Delta}_a- aq^a\Delta_a \nonumber\\ &&+a\left(\sigma^{ab}+\omega^{ab}\right)\Delta_aA_b\,,  \label{A2}
\end{eqnarray}
since $R_{ab}=R^c{}_{acb}$ is the Ricci tensor and $q_a=-h_a{}^bR_{ac}u^c$ is the energy flux vector~\cite{TCM,EMM}. The latter vanishes when matter has the form of a perfect fluid and the same also holds for the 4-acceleration ($A_a$) when the pressure is zero (see Eq.~(\ref{lmdcl}) in \S~\ref{ssLDIs}). Then, if we ignore the effects of vorticity, the above reduces to
\begin{equation}
a{\rm D}^a\dot{\Delta}_a= \dot{\mathcal{D}}+ \sigma^{ab}\mathcal{D}_{\langle ab\rangle}\,.  \label{A3}
\end{equation}
In an exactly analogous way (and under the same conditions) we obtain the nonlinear commutation law $a{\rm D}^a\dot{Z}_a= \dot{\mathcal{Z}}+\sigma^{ab}\mathcal{Z}_{\langle ab\rangle}$.

\section{Alternative derivation of the
%%%%%%%%%%%%%%%%%%%%%%%%%%%%%%%%%%%%%%
$\ddot{\mathcal{D}}$-equation}\label{sA2}
%%%%%%%%%%%%%%%%%%%%%%%%%%%%%%%%%%%%%%%%%
In \S~\ref{ssS-OI} we derived the differential equation of the density contrast ($\mathcal{D}$), namely Eq.~(\ref{2ddotcD1}), after introducing the Einstein-de Sitter background. This allowed us to obtain the 2nd-order limit of (\ref{nldotDelta-a2}) and (\ref{nldotZ-a2}), before taking the comoving 3-divergences of these expressions. As stated in footnote~5, here we will do the reverse. We will first take the nonlinear spatial divergences of Eqs.~(\ref{nldotDelta-a2}), (\ref{nldotZ-a2}) and then obtain the 2nd-order limits of the resulting expressions.

Assuming zero cosmological constant and neglecting any cosmic vorticity, the comoving 3-divergences of Eqs.~(\ref{nldotDelta-a2}) and (\ref{nldotZ-a2}) lead to the nonlinear expressions
\begin{equation}
a{\rm D}^a\dot{\Delta}_a= -\mathcal{Z}- a\Delta^b{\rm D}^a\sigma_{ab}- \sigma_{ab}\mathcal{D}^{\langle ab\rangle}\,,  \label{B1}
\end{equation}
and
\begin{equation}
a{\rm D}^a\dot{Z}_a= -{2\over3}\,\Theta\mathcal{Z}- {1\over2}\,\rho\mathcal{D}- {2\over3}\,Z_aZ^a- {1\over2}\,\rho\Delta_a\Delta^a- aZ^b{\rm D}^a\sigma_{ab}- \sigma_{ab}\mathcal{Z}^{\langle ab\rangle}- 2a^2{\rm D}^2\sigma^2\,,  \label{B2}
\end{equation}
respectively. Introducing an Einstein-de Sitter background will allow us to take the 2nd-order limits of these equations. In doing so, we will only account for the divergence-free part of the linear shear perturbation (i.e.~set ${\rm D}^b\tilde{\sigma}_{ab}=0$ to first order). Then, using the auxiliary second-order commutation laws $a{\rm D}^a\dot{\Delta}_a=\dot{\mathcal{D}}+ \sigma_{ab}\mathcal{D}^{\langle ab\rangle}$ and $a{\rm D}^a\dot{Z}_a=\dot{\mathcal{Z}}+ \sigma_{ab}\mathcal{Z}^{\langle ab\rangle}$ (see Appendix~A), the system (\ref{B1}) and (\ref{B2}) reduces to
\begin{equation}
\dot{\mathcal{D}}+ \mathcal{Z}= -2\tilde{\sigma}_{ab}\tilde{\mathcal{D}}^{\langle ab\rangle}  \label{B3}
\end{equation}
and
\begin{equation}
\dot{\mathcal{Z}}+ 2H\mathcal{Z}+ {1\over2}\,\bar{\rho}\,\mathcal{D}= -{2\over3}\,\tilde{Z}_a\tilde{Z}^a- {1\over2}\,\bar{\rho}\tilde{\Delta}_a\tilde{\Delta}^a- 2\tilde{\sigma}_{ab}\tilde{\mathcal{Z}}^{\langle ab\rangle}\,,  \label{B4}
\end{equation}
on super-Hubble scales (where the shear Laplacian is negligible). We note that the scalars $\mathcal{D}$ and $\mathcal{Z}$ on the left-hand side of (\ref{B3}) and (\ref{B4}) are second-order perturbations, which makes $H$ and $\bar{\rho}$ zero-order variables (see also \S~\ref{ssS-OI} earlier). In analogy, the quantities $\tilde{\Delta}_a$, $\tilde{Z}_a$, $\tilde{\mathcal{D}}_{\langle ab\rangle}$, $\tilde{\mathcal{Z}}_{\langle ab\rangle}$ and $\tilde{\sigma}_{ab}$ have perturbative order one. Comparing (\ref{B4}) to Eq.~(\ref{2dotZ}) in \S~\ref{ssS-OI}, we notice that the different derivation scheme has already added two extra terms to its right-hand side. In what follows, we will consider the implications of these additional terms.

Taking the time derivative of Eq.~(\ref{B3}) and using (\ref{B4}), together with the linear relations (\ref{ldotsh}), (\ref{lshcon2}) and (\ref{ld2}a), gives
\begin{equation}
\ddot{\mathcal{D}}+ 2H\dot{\mathcal{D}}- {1\over2}\,\bar{\rho}\,\mathcal{D}= {1\over2}\,\bar{\rho}\tilde{\Delta}_a\tilde{\Delta}^a+ {2\over3}\,\tilde{Z}_a\tilde{Z}^a+ 2H\tilde{\sigma}_{ab}\tilde{\mathcal{D}}^{\langle ab\rangle}+ 4\tilde{\sigma}_{ab}\tilde{\mathcal{Z}}^{\langle ab\rangle}+ 2\tilde{\mathcal{R}}_{\langle ab\rangle}\tilde{\mathcal{D}}^{\langle ab\rangle}\,,  \label{B5}
\end{equation}
where $\tilde{\mathcal{R}}_{\langle ab\rangle}$ is the linear 3-Ricci tensor. To second order $\tilde{\Delta}_a\tilde{\Delta}^a\propto t^{4/3}$, $\tilde{Z}_a\tilde{Z}^a\propto t^{-2/3}$, $\tilde{\sigma}_{ab}\tilde{\mathcal{D}}^{\langle ab\rangle}\propto t^{1/3}$, $\tilde{\sigma}_{ab}\tilde{\mathcal{Z}}^{\langle ab\rangle}\propto t^{-2/3}$ and $\tilde{\mathcal{R}}_{\langle ab\rangle}\tilde{\mathcal{D}}^{\langle ab\rangle}\propto t^{-2/3}$ (see \S~\ref{ssRST}, \S~\ref{ssLGWs} and \S~\ref{ssLDIs}), while $H=2/3t$ and $\bar{\rho}=4/3t^2$  to zero order. Employing these evolution laws, Eq.~(\ref{B5}) recasts into
\begin{equation}
3t^2\ddot{\mathcal{D}}+ 4t\,\dot{\mathcal{D}}- 2\mathcal{D}= 2\alpha_0\left({t\over t_0}\right)^{4/3}\,.  \label{B6}
\end{equation}
This expression is (formalistically) identical to Eq.~(\ref{2ddotcD2}) obtained in \S~\ref{ssS-OI} and therefore accepts the same solution, namely
\begin{eqnarray}
\mathcal{D}&=& {2\over5}\left(\mathcal{D}_0 -{\dot{\mathcal{D}}_0\over H_0}+{3\over7}\,\alpha_0\right) \left({t_0\over t}\right)+ {3\over5}\left(\mathcal{D}_0+{2\over3}\,{\dot{\mathcal{D}}_0\over H_0}- \alpha_0\right)\left({t\over t_0}\right)^{2/3} \nonumber\\ &&+{3\over7}\,\alpha_0\left({t\over t_0}\right)^{4/3}\,.  \label{B7}
\end{eqnarray}
The only difference between the above and solution (\ref{2ndcD2}) is in the dimensionless coefficient $\alpha_0$, which carries the combined effect of the 2nd-order source terms. Here, this factor is no longer given by (\ref{alpha02}) but by
\begin{equation}
\alpha_0= \tilde{\Delta}_0^2+ {4\over9}\left({\tilde{Z}_0\over H_0}\right)^2+ {4\over3}\,{\tilde{\sigma}_0\over H_0} \tilde{\mathcal{D}}_0+ {8\over3}\,{\tilde{\sigma}_0\over H_0}{\tilde{\mathcal{Z}}_0\over H_0}+ {4\over3}\,{\tilde{\mathcal{R}}_0\over H_0^2}\tilde{\mathcal{D}}_0\,.  \label{B8}
\end{equation}
As in Eq.~(\ref{alpha02}), the linear quantities $\tilde{\Delta}_0$, $\tilde{Z}_0$, $\tilde{\mathcal{D}}_0$, $\tilde{\mathcal{Z}}_0$, $\tilde{\sigma}_0$ and $\tilde{\mathcal{R}}_0$ have been evaluated at recombination (i.e.~$t_0=t_{RC}$), but they can generally vary in space. In contrast to (\ref{alpha02}), there are two additional terms on the right-hand side of the above (the first two). Nevertheless, since $\tilde{\Delta}_0$, $\tilde{Z}_0/H_0\ll1$, the aforementioned extra source-terms do not alter the overall effect of the gravitational waves, as described in \S~\ref{ssIs}.\footnote{Adding 2nd-order source terms to the right-hand side of Eq.~(\ref{B6}) can only introduce faster growing modes to the associated solution. Indeed, extra source terms that evolve slower than $t^{4/3}$ will quickly fade away. Those evolving faster, on the other hand, could further accelerate the growth of $\mathcal{D}$ (if/when they dominate).} The Weyl field could still drive scalar density perturbations to a considerably faster growth-rate, but only at very late times.

\end{document}